\documentclass[reprint,aps,prd,nofootinbib,preprintnumbers,twocolumns,showpacs,10pt, superscriptaddress]{revtex4-1}
\usepackage{epsfig}
\usepackage{epstopdf}
\usepackage{hyperref}
\usepackage{amssymb}
\usepackage{amsmath}
\usepackage{amsfonts}
\usepackage{graphicx}
\usepackage{bm}
\usepackage[usenames,dvipsnames]{color}
\hypersetup{
  colorlinks   = true, %Colours links instead of ugly boxes
  urlcolor     = blue, %Colour for external hyperlinks
  linkcolor    = blue, %Colour of internal links
  citecolor    = red   %Colour of citations
}
\def	\be	{\begin{equation}}
\def	\ee	{\end{equation}}
\def	\bqt	{\begin{quote}}
\def	\eqt	{\end{quote}}

\begin{document}

\title{Non-linear energy conservation theorem in the framework of Special Relativity}

\author{Gin\'{e}s R.P\'{e}rez Teruel}

\affiliation{$^1$Departamento de F\'{i}sica Te\'{o}rica, Universidad de Valencia, Burjassot-46100, Valencia, Spain} 

\begin{abstract}
\begin{center}
{\bf Abstract}
\end{center}
\noindent
In this work we revisit the study of the gravitational interaction in the context of the Special Theory of Relativity. It is found that, as long as the equivalence principle is respected, a relativistic non-linear energy conservation theorem arises in a natural way. We interpret that this non-linear conservation law stresses the non-linear character of the gravitational interaction.The theorem reproduces the energy conservation theorem of Newtonian mechanics in the corresponding low energy limit, but also allows to derive some standard results of post-Newtonian gravity, such as the formula of the gravitational redshift. Guided by this conservation law, we develop a Lagrangian formalism for a particle in a gravitational field. We realize that the Lagrangian can be written in an explicit covariant fashion, and turns out to be the geodesic Lagrangian of a curved Lorentzian manifold. Therefore, any attempt to describe gravity within the Special Theory, leads outside their own domains towards a curved space-time. Thus, the pedagogical content of the paper may be useful as a starting point to discuss the problem of Gravitation in the context of the Special Theory, as a preliminary step before introducing General Relativity.
\end{abstract}

\maketitle
\section{Introduction}
\label{Introduction} 
\thispagestyle{empty}

\noindent
The General Theory of Relativity (GR) is the most accepted theory nowadays to describe the behaviour of the classical gravitational field. The theory is probably one of the most well-tested theories in physics. In general, there has been an excellent level of agreement between theory and experiments in scales that range from millimeters to astronomical units, scales in which weak and strong field phenomena can be observed \cite{Will-LR}. Due to their accurate predictions, it is generally accepted that the theory should also work at larger and shorter scales, and at weaker and stronger regimes. \\

Nevertheless, there are serious and fundamental open problems that remain to be solved. For instance, to explain the rotation curves of spiral galaxies, we must accept the existence of vast amounts of unseen matter surrounding those galaxies.  A similar situation occurs with the analysis of the light emitted by distant type-Ia supernovae and some properties of the  distribution of matter and radiation at large scales. To make sense of those observations within the framework of GR, we must accept the existence of yet another source of energy with repulsive gravitational properties\cite{reviews}. In addition, the outstanding difficulties to harmonize the conceptual and mathematical framework of GR with the rest of physics, in particular the problem of consistently combining GR with Quantum Mechanics (QM), have risen the interest of the physics community in the search for modified theories of gravity. These modified theories of gravity include a wide variety of different approximations: Mond theories \cite{MOND}\cite{Milgrom:2009ee}, scalar-tensor theories\cite{Chiba}, $f(R)$ generalizations in metric and Palatini formalism respectively \cite{$f(R)$,Ter,Teru}, and even studies about the implications of a violation of the weak and strong equivalence principle. (see Ref.\cite{Bousso} for a recent one). Due to the fact that the rest of the classical and quantum field theories are formulated in a flat spacetime, a natural possibility seems the seek for a relativistic theory of gravitation purely constructed in Minkowski spacetime, i.e, a theory of gravity subjected only to the constraints and principles of Special Relativity (SR). Indeed, this approach was historically the first considered by Einstein himself, although he eventually abandoned it to pursue a theory constructed with the aid of Riemannian geometry. The reasons to understand the rejection of such a promising approach are complex and sometimes not well explained in the introductory lectures of GR, where the theory is presented as a sort of revelation. Even in our days, we still see theoretical attempts to search for a theory of gravitation in Minkowski's spacetime, with the hope that such achievement will make more factible the unification of gravity with the rest of physics. Then, why is this a failed research program? Why is not possible to find a satisfactory special-relativistic theory of gravitation without crossing the own domains of SR?

In this work, we revisit the problem of gravitation in SR with open mind , and find that if we assume that the equivalence principle is of universal validity, then it can be derived a non-linear conservation law that relates the variation of the relativistic mass with the variation of the gravitational potential. This formula generalizes in a natural way the energy conservation theorem of Newtonian Mechanics. Furthermore, the Lagrangian formalism that we obtain for the model possesses an intrinsic gometric meaning: it can be rewritten as the geodesic Lagrangian of a Lorentzian manifold. Then, any consistent attempt to describe gravity in the framework of SR, will lead outside the domains of this theory. We conclude that our approach, due to their simplicity, may be employed in a pedagogical manner to introduce the problem of Gravitation in SR as a previous step before addressing the study of GR, and it can provide a tool to understand why gravity is different from the other forces of nature: Gravitation cannot be described by any flat space-time-based theory.
\section{The equivalence principle and the non-linear energy conservation theorem}
\label{Derivation} 
\thispagestyle{empty}

\noindent
Let us consider the movement of a test particle in an external gravitation field in the framework of SR. We assume that the four-momentum $p^{\mu}$, of the test particle in an arbitrary reference frame is given by $(E,c\vec{p})$. Any differential variation of their energy due to their movement can be expressed as
\be
dE=dm_{i}c^{2}
\ee
where $dm_{i}$ denotes the differential variation of the inertial relativistic mass of the test particle. On the other hand, since the particle is subjected to a conservative force that derives from a potential, the infinitesimal work made by the gravitational field on the particle will be
\be
dW=-m_{g}\nabla\phi \cdot dr=-m_{g}d\phi
\ee

Where $m_{g}$ is the gravitational mass of the particle. Equaling both equations, we get
\be
dm_{i}c^{2}=-m_{g}d\phi
\ee
The acceptance of the equivalence principle implies that the massess that appear in both sides of the last equation are strictly equal. In these conditions, if $m_{i}=m_{g}$ we can rearrange terms to write
\be
\int_{r_{1}}^{r_{2}}\frac{dm}{m}=-\int_{r_{1}}^{r_{2}}\frac{d\phi}{c^{2}}
\ee

The integration of this equation between two arbitrary points provides, 
\be\label{mass-potential formula}
m_{1}(\phi)e^{\phi_{1}/c^{2}}=m_{2}(\phi)e^{\phi_{2}/c^{2}}
\ee

Or equivalently,
\be\label{conservation law}
\frac{m_{0}}{\sqrt{1-\beta^{2}(\phi)}}\exp(\phi/c^2)=C
\ee
Where $C$ is a constant of motion, and $m_{0}$ is a constant characteristic of the particle which does not depend on $\phi$. What is the physical meaning of this formula? It represents a non-linear generalization of the energy conservation theorem of newtonian mechanics. Note that this conservation law emerges automatically as long as the equivalence principle is accepted. In plain words, equation (\ref{mass-potential formula}) is expressing the following: when the particle moves far from the sources of the gravitational field, the value of the potential $\phi$ increases, and therefore their relativistic mass, $m(\phi)=\frac{m_{0}}{\sqrt{1-\beta^{2}(\phi)}}$ should decrease because the factor, $\beta$ decreases. Note that the opposite situation takes place when the particle approximates to the sources of the gravitational field: in this case the potential $\phi$ decreases (becomes more negative), and then the relativistic mass increases due to the increase of the velocity of the particle. We should mention that the exponential mathematical form of the function $m(\phi)$ has been obtained elsewhere (see for instance \cite{Vankov},\cite{Ben-Amots}) although the presence of a general conservation law is not recognized in these works, nor their full implications. Indeed, we will show in the next sections that guided by this conservation theorem it is possible to formulate a Lagrangian theory that can be interpreted in geometrical terms as the geodesic Lagrangian of a Lorentzian manifold.
\subsection{The Newtonian limit}
\label{classical} 
\thispagestyle{empty}

\noindent
In order to prove the robustness of this result, let us consider the case of a particle initially placed at rest at a point $r_{0}$ from the origin of a coordinate system. Suppose that the measured value of the potential at $r_{0}$ is $\phi_{0}$. If the particle moves to another point $r_{1}$ where the potential is $\phi_{1}$, under the effect of the gravitational field, their relativistic mass at $r_{1}$ according to Eqs. (\ref{mass-potential formula}-\ref{conservation law}) will be:
\begin{equation}
m_{1}(\phi)=m_{0}e^{(\phi_{0}-\phi_{1})/c^{2}}
\end{equation}

Then, according to STR the energy of the particle measured by the same observer at the point $r_{1}$ will be
\begin{equation}\label{energy-balance}
E_{1}=m_{1}(\phi)c^{2}=m_{0}c^{2}e^{(\phi_{0}-\phi_{1})/c^{2}}=T_{1}+m_{0}c^{2}
\end{equation}
where $T_{1}$ is the kinetic energy acquired by the particle due to the value of their velocity at $r_{1}$. Note that if the variation, $\phi_{0}-\phi_{1}$ is small compared with $c^{2}$, we can approximate Eq. (\ref{energy-balance}) as follows

\begin{align}
\displaystyle E_{1}&=m_{0}c^{2}e^{(\phi_{0}-\phi_{1})/c^{2}}= m_{0}c^{2}\Big(1+(\phi_{0}-\phi_{1})/c^{2}+...\Big)\nonumber\\
&=\displaystyle m_{0}c^{2}+m_{0}(\phi_{0}-\phi_{1})=T_{1}+m_{0}c^{2}
\,
\end{align}
This implies,
\begin{equation}\label{E-conservation}
\Delta T= T_{1}-T_{0}=m_{0}(\phi_{0}-\phi_{1})=-\Delta U
\end{equation}
This result recovers in a natural way the conservation of the mechanical energy of newtonian mechanics. Indeed, note that the particle was initially at rest, ($T_{0}=0$) therefore Eq.(\ref{E-conservation}) express that the increase of the kinetic energy $\Delta T$ of the particle is equal to the decrease of their potential energy. On the other hand, using the general energy-momentum relation $E^{2}=c^{2}p^{2}+m_{0}^{2}c^{4}$, we can find a closed formula for the momentum of the particle at the point $r_{1}$ in terms of the variation of the potential between both points as follows
\begin{equation}
p=\frac{1}{c}\sqrt{E^{2}-m_{0}^{2}c^{4}}=m_{0}c\sqrt{e^{2(\phi_{0}-\phi_{1})/c^{2}}-1}
\end{equation}
This relation is completely general. However, when $(\phi_{0}-\phi_{1})/c^{2}<<1$, the approximation of the exponential up to the second order provides: 

\begin{align}
p_{1}&\simeq m_{0}c\sqrt{1+2(\phi_{0}-\phi_{1})/c^{2}-1}=m_{0}c\sqrt{2(\phi_{0}-\phi_{1})/c^{2}}\nonumber\\
&=\sqrt{2m_{0}T_{1}}
\end{align}
\\
In total agreement with the result of non-relativistic classical mechanics, i.e, $T=p^{2}/2m$

\subsection{The linear energy conservation theorem of classical mechanics as a particular case}
\label{nonlinear} 
\thispagestyle{empty}

\noindent
We can provide a more transparent proof of how the non-linear energy conservation theorem (\ref{mass-potential formula}), reduces to the linear conservation of energy of Newtonian mechanics, $E=T+U$ in the appropriate limit. Indeed, note that (\ref{mass-potential formula}), can be written in the form\\
 
\begin{equation}
\frac{m_{0}}{\sqrt{1-\beta^{2}_{1}}}e^{\phi_{1}/c^{2}}=\frac{m_{0}}{\sqrt{1-\beta^{2}_{2}}}e^{\phi_{2}/c^{2}}
\end{equation}

For small velocities compared with the speed of light, $\beta^{2}=v^{2}/c^{2}<<1$, and weak gravitational fields, $\phi/c^{2}<<1$, the last equation can be approximated as
\begin{equation}
m_{0}\Big(1+\frac{\beta^{2}_{1}}{2}\Big)\Big(1+\frac{\phi_{1}}{c^{2}}\Big)=m_{0}\Big(1+\frac{\beta^{2}_{2}}{2}\Big)\Big(1+\frac{\phi_{2}}{c^{2}}\Big)
\end{equation}
\\
where we have neglected the contributions that go as $\mathcal{O}(1/c^{4})$. After a direct computation, the last equation provides the result:
\begin{equation}
\frac{1}{2}m_{0}v^{2}_{1}+m_{0}\phi_{1}+\frac{1}{2}m_{0}\frac{v^{2}_{1}\phi_{1}}{c^{2}}=\frac{1}{2}m_{0}v^{2}_{2}+m_{0}\phi_{2}+\frac{1}{2}m_{0}\frac{v^{2}_{2}\phi_{2}}{c^{2}}
\end{equation}
\\

As we can see, this is the conservation theorem of the mechanical energy of classical physics, $T+U=C$, corrected by a first non-linear contribution given by, $m_{0}v^{2}\phi/2c^{2}$. It is interesting to note that the presence of a non-linear energy conservation theorem stresses the non-linear character of the gravitational interaction. We will show in the next section that the full non-linear conservation law (\ref{mass-potential formula}-\ref{conservation law}) can be viewed as a conserved canonical Hamiltonian.
\subsection{Lagrangian formulation and equations of motion}
We postulate the following non-linear relativistic Lagrangian for a particle in a gravitational field:
\\
\begin{equation}\label{Lagrangian}
\mathcal{L}=-m_{0}c^{2}\sqrt{1-\beta^{2}}\exp(\phi/c^{2})
\end{equation}
\\
In the non-relativistic limit this Lagrangian becomes:
\begin{equation}
\mathcal{L}\approx -m_{0}c^{2}+\frac{1}{2}m_{0}v^{2}-m_{0}\phi+\frac{1}{2}m_{0}\phi\frac{v^{2}}{c^{2}}+...\approx T-U-m_{0}c^{2}
\end{equation}
Then, this Lagrangian possesses an acceptable low energy behavior. Now, in order to proceed further, we assume that the relation between the Lagrangian $\mathcal{L}$, and the Hamiltonian $\mathcal{H}$ is the standard, i.e, they are related through a Legendre transform, namely: 
\begin{equation}\label{Hamiltonian}
\mathcal{H}=\sum_{i}p_{i}v_{i}-\mathcal{L}
\end{equation}
Where the canonical momenta $p_{i}$, are given by,
\begin{equation}
p_{i}= \frac{\partial \mathcal{L}}{\partial v_{i}}=\frac{m_{0}v_{i}}{\sqrt{1-\beta^{2}}}\exp(\phi/c^{2})
\end{equation}
And we have assumed that the potential $\phi(r)$ is only a function of $r$ and does not depend on the velocities.
Substituing this result in (\ref{Hamiltonian}) we obtain
\begin{align}\label{Generalized_Hamiltonian}
\mathcal{H}&=\Big(\sum_{i}\frac{m_{0}v_{i}^{2}}{\sqrt{1-\beta^{2}}}+m_{0}c^{2}\sqrt{1-\beta^{2}}\Big)\exp(\phi/c^{2})\nonumber\\
&=\Big(\frac{\sum_{i}m_{0}v_{i}^{2}+m_{0}c^{2}(1-\beta^{2})}{\sqrt{1-\beta^{2}}}\Big)\exp(\phi/c^{2})\nonumber\\
&=\frac{m_{0}c^{2}}{\sqrt{1-\beta^{2}}}\exp(\phi/c^{2})
\end{align}
Therefore, as we suspected, the non-linear conservation law (\ref{mass-potential formula},\ref{conservation law}), is nothing but the conservation of the canonical Hamiltonian (\ref{Generalized_Hamiltonian}), times a constant.
On the other hand, the equations of motion for the Kepler problem can be obtained by application of the Euler-Lagrange equations to the Lagrangian (\ref{Lagrangian}). These equations for the case of movement in two dimensions will be:

\begin{equation}
  \label{eq:euler-lag-likeEQS}
  \frac{\partial \mathcal{L}}{\partial r}-\frac{d}{dt}\frac{\partial \mathcal{L}}
  {\partial \dot r}=0\,
\end{equation}
\begin{equation}
  \label{eq:euler-lag-likeEQS}
  \frac{\partial \mathcal{L}}{\partial \theta}-\frac{d}{dt}\frac{\partial \mathcal{L}}
  {\partial \dot \theta}=0\,
\end{equation}
Since, $\beta^{2}=v^{2}/c^{2}=({\dot r^{2}}+r^{2}{\dot \theta}^{2})/c^{2}$, the Lagrangian function (\ref{Lagrangian}) will acquire the following form
\begin{equation}\label{Lagrangian2}
\mathcal{L}=-m_{0}c^{2}\sqrt{1-\frac{{\dot r^{2}}+r^{2}{\dot \theta}^{2}}{c^{2}}}\exp(\phi(r)/c^{2})
\end{equation}

Then, the coordinate $\theta$ is cyclic, which implies that a conserved constant of motion is present. Indeed, applying (\ref{eq:euler-lag-likeEQS}) we obtain
\begin{equation}\label{conservation_l}
\frac{d}{dt}\Big(\frac{m_{0}r^{2}\dot \theta}{\sqrt{1-\beta^{2}}}\exp(\phi(r)/c^{2})\Big)=\frac{d}{dt}(\mathcal{H}r^{2}\dot \theta)=0
\end{equation}
Since we know that, $\mathcal{H}=\exp(\phi/c^{2})m_{0}/\sqrt{1-\beta^{2}}$ is constant given the previous results, it is easy to see that equation (\ref{conservation_l}) is expressing the conservation of the angular momentum, $r^{2}\dot \theta=C$. Regarding the other Euler-Lagrange equation, with a bit of algebra we find:
\small
\begin{equation}
\Big(-\frac{\partial\phi}{\partial r}m_{0}\sqrt{1-\beta^{2}}+\frac{l^{2}}{m_{0}r^{3}\sqrt{1-\beta^{2}}}\Big) e^{\phi/c^{2}}=\frac{d}{dt}\Big(\frac{m_{0}\dot r}{\sqrt{1-\beta^{2}}}e^{\phi/c^{2}}\Big)
\end{equation}
\normalsize
Where, $l=m_{0}r^{2}{\dot \theta}^{2}$. According to these equations of motion, the product $m_{0}\cdot\exp(\phi/c^{2}$), determines the inertia of a mass point. In other words: the bigger the potential is, the bigger will be the resistence exerted by the particle in response to a variation in its velocity. Developing the derivative of the right hand side, we can remove the factor $e^{\phi/c^{2}}$ from the analysis. Doing this, we obtain the compact form of the equation of motion for the Kepler problem as:
\begin{align}
\frac{d}{dt}\Big(\frac{m_{0}\dot r}{\sqrt{1-\beta^{2}}}\Big)=&-\frac{\partial\phi}{\partial r}m_{0}\sqrt{1-\beta^{2}}+\frac{l^{2}}{m_{0}r^{3}\sqrt{1-\beta^{2}}}\nonumber\\
&-\frac{m_{0}\dot r}{c^{2}\sqrt{1-\beta^{2}}}\frac{d\phi}{dt}
\end{align}
The right hand side can be seen as the effective force that feels the particle. When $\beta<<1$, we naturally recover the equation of motion of Newtonian mechanics. However, the situation is more involved when high velocities are considered, because the term of the right hand side that includes the gradient decreases (multiplies the factor $\sqrt{1-\beta^{2}}$), while the other two increase with the velocity. This effect may have interesting consequences, for example, to compute the prediction of this model for the anomalous perihelion precession of planetary orbits; Indeed, the comparison between this prediction with those of GR, will falsify the model, or at least will fix strong constraints on the structure of the family of possible non-linear Lagrangians.
On the other hand, the non-linear Lagrangian (\ref{Lagrangian}), possesses a deep geometrical meaning. Indeed, note that it can be rewritten as:
\begin{align}\label{Lagrangian3}
\mathcal{L}&=-m_{0}c^{2}\sqrt{1-\beta^{2}}e^{\phi/c^{2}}=-m_{0}c^{2}\sqrt{\exp(2\phi/c^{2})(1-\beta^{2})}\nonumber\\
&=-\frac{m_{0}c}{dt}\sqrt{g_{\mu\nu}(x)dx^{\mu}dx^{\nu}}=-m_{0}c\frac{ds}{dt}
\end{align}

Where, $g_{\mu\nu}(x)=\exp(2\phi/c^{2})\cdot\eta_{\mu\nu}$. This is the geodesic Lagrangian of a curved manifold! Then, if we try to formulate the gravitational interaction within a flat space-time, the consistency of the theory will push you out of the Special Theory towards a curved space-time. Therefore, the Special Theory of Relativity cannot contain by itself the gravitaional interaction. This can be useful to explain why gravitation is different from the other fundamental forces of nature.\\ 

\section{Summary and conclusions}
In this work, we have revisited the problem of the movement of a particle in a gravitational field in the context of SR. We have found that the equality of inertial and gravitational massess allows to derive a non-linear energy conservation law that generalizes the energy conservation theorem of Newtonian mechanics. This non-linear conservation law has provided a hint of how to construct a consistent Lagrangian formalism. As we expected, the Lagrangian function turns out to be non-linear, which via the corresponding Euler-Lagrange equations, provides equations of motion that are also non-linear, but with a correct classical limit. In addition, we have realized that the Lagrangian function possesses a deep geometrical insight. In fact, it can be rewritten as the covariant geodesic Lagrangian of a curved manifold. Then, starting from the Special Theory perspective, the consistency of the theory implies a natural transition from a flat space-time to a curved space-time. This can justify the pedagogical use of this model to illustrate a beautiful transition between SR and GR. On the other hand, it would be very interesting for future works the study of possible generalizations of this conservation law for velocity-dependent potentials of Berger's type \cite{Berger}, or even for more general retarded potentials\cite{Gine}, which have been shown to reproduce the anomalous precession of Mercury's perihelion.
\section{Acknowledgements}
The author would like to thank the anonymous referee for helpful comments and suggestions that improved the quality of the manuscript.

\end{document}